\documentstyle[12pt]{article}
\catcode`\@=11

\@addtoreset{equation}{section}
\def\theequation{\arabic{section}.\arabic{equation}}

\catcode`\@=11

\def\section{\@startsection{section}{1}{\z@}{3.5ex plus 1ex minus
   .2ex}{2.3ex plus .2ex}{\large\bf}}

\long\def\@makefntext#1{\parindent 0cm\noindent
\hbox to 1em{\hss$^{\@thefnmark}$}#1}

\def\eqnarray{\let\@currentlabel=\theequation\refstepcounter{equation}
    \global\@eqnswtrue
    \global\@eqcnt\z@\tabskip\@centering\let\\=\@eqncr
    $$\halign to \displaywidth\bgroup\@eqnsel\hskip\@centering
      $\displaystyle\tabskip\z@{##}$&\global\@eqcnt\@ne
       \hfil${{}##{}}$\hfil
      &\global\@eqcnt\tw@ $\displaystyle\tabskip\z@{##}$\hfil
       \tabskip\@centering&\llap{##}\tabskip\z@\cr}
\def\lefteqn#1{\hbox to 4\arraycolsep{$\displaystyle #1$\hss}}
\begin{document}
\begin{titlepage}
\vspace{.5in}
\begin{flushright}
UCD-98-18\\
hep-th/9812013\\
November 1998\\
revised January 1999\\
\end{flushright}
\vspace{.5in}
\begin{center}
{\Large\bf
Black Hole Entropy\\[1ex] from Conformal Field Theory\\[1.2ex]
in Any Dimension}\\
\vspace{.4in}
{S.~C{\sc arlip}\footnote{\it email: carlip@dirac.ucdavis.edu}\\
       {\small\it Department of Physics}\\
       {\small\it University of California}\\
       {\small\it Davis, CA 95616}\\{\small\it USA}}
\end{center}

\vspace{.5in}
\begin{center}
{\large\bf Abstract}
\end{center}
\begin{center}
\begin{minipage}{5.2in}
{\small
Restricted to a black hole horizon, the ``gauge'' algebra of surface 
deformations in general relativity contains a Virasoro subalgebra 
with a calculable central charge.  The fields in any quantum theory 
of gravity must transform accordingly, i.e., they must admit a conformal 
field theory description.  Applying Cardy's formula for the asymptotic 
density of states, I use this result to derive the Bekenstein-Hawking 
entropy.  This method is universal---it holds for any black hole, and 
requires no details of quantum gravity---but it is also explicitly 
statistical mechanical, based on counting microscopic states.
}
\end{minipage}
\end{center}
\end{titlepage}
\addtocounter{footnote}{-1}

\section{Introduction}

Since the discovery that black holes behave as thermal objects, an 
outstanding open question has been whether black hole thermodynamics 
has a ``statistical mechanical'' description in terms of microscopic 
states.  An answer could shed light on broader problems of quantum 
gravity; at a minimum, black hole thermodynamics provides an important 
consistency check for any candidate theory of quantum gravity.

Until recently, we had no convincing model of microscopic black hole 
states.  Today, we have a plethora of possibilities, from D-brane states 
in string theory \cite{Dbrane} to spin network states in loop quantum 
gravity \cite{Ashtekar}.  A fundamental issue remains, however, perhaps 
best described as the problem of universality \cite{Strominger1}.  The 
Bekenstein-Hawking entropy can be computed entirely within the framework 
of quantum field theory in a fixed curved background.  It is hard to
see how such a calculation could ``know'' the details of a microscopic 
gravitational theory.  Rather, it seems more likely that some unknown
universal mechanism forces {\em any\/} suitable quantum theory to give 
the standard result.

A major step toward finding a universal mechanism was taken by Strominger 
\cite{Strominger2}, who reanalyzed the (2+1)-dimensional black hole 
\cite{BTZ} using conformal field theory methods.  Brown and Henneaux 
had shown that the asymptotic symmetry algebra for this solution was a 
Virasoro algebra, implying that any theory of microscopic states should 
be a conformal field theory \cite{Brown}.  Strominger observed that the 
Cardy formula \cite{Cardy} for the asymptotic growth of states could thus 
be used to compute the entropy, and that the result agreed with the usual 
Bekenstein-Hawking expression.  This analysis was subsequently extended 
to a number of higher-dimensional black holes with near-horizon geometries 
resembling that of the (2+1)-dimensional black hole (see \cite{Carlip1} 
for a partial list of references).

But while many black holes have the appropriate near-horizon geometry
for such an analysis, others do not.  Moreover, the Virasoro algebra of
Brown and Henneaux is an algebra of asymptotic symmetries at spatial
infinity, while black hole entropy should arguably be a more local
property of horizons.

In this paper, I generalize Strominger's approach by looking at the
symmetries of the horizon of an arbitrary black hole.  The relevant
algebra of surface deformations contains a physically important Virasoro
algebra, essentially consisting of deformations of the $r$--$t$ plane
that leave the horizon fixed.  The analysis of Brown and Henneaux can
be extended to find the central charge of this algebra.  I show that
the Cardy formula then yields the correct Bekenstein-Hawking entropy,
independent of the details of the black hole.

\section{Metric and Boundary Terms}

Let us start with a general black-hole-like metric in $n$ spacetime
dimensions,\footnote{Greek letters from the middle of the alphabet are
spacetime indices, Roman letters are spatial indices, and Greek letters
from the beginning of the alphabet are ``boundary'' or ``angular'' indices.}
\begin{equation}
ds^2 = -N^2 dt^2 + f^2(dr + N^r dt)^2
     + \sigma_{\alpha\beta}(dx^\alpha + N^\alpha dt)(dx^\beta + N^\beta dt) ,
\label{a1}
\end{equation}
with a lapse function $N$ that vanishes at a horizon $r=r_+$ and behaves
near $r=r_+$ as
\begin{equation}
N^2 = h(x^\alpha)(r-r_+) + O(r-r_+)^2 , \qquad
n^a\partial_a N = {2\pi /\beta} ,
\label{a2}
\end{equation}
where $n^a$ is the unit normal to $r=r_+$ on a constant $t$ slice.  For a 
stationary black hole with coordinates such that $N^r=0$, $\beta$ is the 
inverse Hawking temperature, and is constant on the horizon.

I will treat $r=r_+$ as a boundary---or, more precisely, a surface at which
certain fields are fixed, and at which boundary terms are therefore needed
in a variational principle \cite{Carlip2}---and will assume that the metric
approaches that of a standard, momentarily stationary, black hole near this
boundary.  Suitable fall-off conditions near $N=0$ are
\begin{eqnarray}
&&f = {\beta h\over4\pi}N^{-1} + O(1) , \qquad N^r = O(N^2) , \qquad
\sigma_{\alpha\beta} = O(1) , \qquad N^\alpha = O(1), \nonumber\\
&&(\partial_t - N^r\partial_r) g_{\mu\nu} = O(N) g_{\mu\nu},
\qquad \nabla_\alpha N_\beta + \nabla_\beta N_\alpha= O(N) .
\label{a3}
\end{eqnarray}
The last condition is essentially the requirement that angular velocity be
constant on the horizon.  The extrinsic curvature of a slice of constant 
time then behaves as
\begin{equation}
K_{rr} = O(1/N^3) , \qquad K_{\alpha r} = O(1/N) ,
\qquad K_{\alpha\beta} = O(1)
\label{a4}
\end{equation}
near the horizon.  (Note that $\partial_r N = O(1/N)$.)

In the Hamiltonian (ADM) formulation of general relativity, the group of 
``gauge'' symmetries is the surface deformation group, generated by the 
quantity
\begin{equation}
H[{\hat\xi}] = \int_\Sigma d^{n-1}x\, {\hat\xi}^\mu{\cal H}_\mu +
\hbox{\em boundary terms} ,
\label{a5}
\end{equation}
where $\{ {\cal H}_t, {\cal H}_a \}$ are the Hamiltonian and momentum
constraints.  The parameters ${\hat\xi}^\mu$ are almost, but not quite, 
identical to parameters $\xi^\mu$ labeling infinitesimal spacetime 
diffeomorphisms; the two are related by \cite{Brown2}
\begin{equation}
{\hat\xi}^t = N\xi^t, \qquad {\hat\xi}^a = \xi^a + N^a\xi^t .
\label{a6}
\end{equation}
As usual, the variation of the ``volume piece'' of $H[{\hat\xi}]$ contains
surface terms at the boundary---in this case, the horizon---which take the
standard form \cite{Brown}
\begin{equation}
-{1\over16\pi G} \int_{r=r_+}\!\! d^{n-2}x\, \left\{
\sqrt{\sigma}\left(\sigma^{ac}n^b - \sigma^{ab}n^c\right)
\left({\hat\xi}^t\nabla_c\delta g_{ab}
     - \nabla_c{\hat\xi}^t\delta g_{ab}\right)
+ 2 {\hat\xi}^a\delta\pi_a{}^r - {\hat\xi}^r \pi^{ab}\delta g_{ab} \right\} ,
\label{a7}
\end{equation}
where $\pi^{ab} = f\sqrt{\sigma}(K^{ab} - g^{ab}K )$ is the momentum 
conjugate to $g_{ab}$.  To have a well-defined symmetry generator, we must 
add a boundary term to $H[{\hat\xi}]$ to cancel the variation (\ref{a7}).  
Note that the fall-off conditions (\ref{a3})--(\ref{a4}) necessitate that
\begin{equation}
{\hat\xi}^r = O(N^2), \qquad  {\hat\xi}^t = O(N),
\qquad {\hat\xi}^\alpha = O(1)  ,
\label{a9}
\end{equation}
since otherwise surface deformations would change the metric near $N=0$.

It is straightforward to check that a term 
\begin{equation}
J[{\hat\xi}] = {1\over8\pi G}\int_{r=r_+}\!\! d^{n-2}x\, \Bigl\{
  n^a\nabla_a{\hat\xi}^t\sqrt{\sigma}
  + {\hat\xi}^a\pi_a{}^r + n_a{\hat\xi}^a K\sqrt{\sigma} \Bigr\}
\label{a10}
\end{equation}
added to the generator (\ref{a5}) yields a variation
\begin{equation}
\delta\left(H[{\hat\xi}] + J[{\hat\xi}]\right) = \hbox{\em bulk terms}
+ {1\over8\pi G} \int_{r=r_+}\!\! d^{n-2}x\, \left(
\delta n^r\partial_r{\hat\xi}^t + {1\over f}{\hat\xi}^r\delta K_{rr}
+ \delta n_r{\hat\xi}^r K\right)\sqrt{\sigma} .
\label{a11}
\end{equation}
In constrast to more familiar variational problems, the normal $n^a$ need
not be fixed at the boundary, but $\delta n^r$ can be computed from the
requirement that $\delta(g_{ab}n^an^b) = 0$.  If we now restrict our
variations to those satisfying $\delta f/ f = O(N)$ and $\delta K_{rr}/ 
K_{rr} = O(N)$, the boundary term in (\ref{a11}) vanishes, as required.

A useful check of eqn.\ (\ref{a10}) can be obtained by specializing to
variations $\delta H$ that are themselves surface deformations.  Let
$L[{\hat\xi}] = H[{\hat\xi}]+J[{\hat\xi}]$ be the full generator of surface 
deformations.  Then the deformation of $L[{\hat\xi}]$ should itself be 
generated by $L[{\hat\xi}]$: that is, it should be given by the Poisson 
bracket \cite{Brown}
\begin{equation}
\delta_{{\hat\xi}_2}L[{\hat\xi}_1] =
\left\{ L[{\hat\xi}_2], L[{\hat\xi}_1] \right\}
= L[\{ {\hat\xi}_1, {\hat\xi}_2 \}_{\hbox{\scriptsize SD}}] +
K[{\hat\xi}_1,{\hat\xi}_2]
\label{a13}
\end{equation}
where $K[{\hat\xi}_1,{\hat\xi}_2]$ is a possible central term in the
algebra.  Here $\{ {\hat\xi}_1, {\hat\xi}_2 \}_{\hbox{\scriptsize SD}}$
is the Lie bracket for the algebra of surface deformations, given by
\cite{Brown2}
\begin{eqnarray}
\{ {\hat\xi}_1, {\hat\xi}_2 \}_{\hbox{\scriptsize SD}}^t &=&
{\hat\xi}_1^a\partial_a{\hat\xi}_2^t - {\hat\xi}_2^a\partial_a{\hat\xi}_1^t
\nonumber\\
\{ {\hat\xi}_1, {\hat\xi}_2 \}_{\hbox{\scriptsize SD}}^a &=&
{\hat\xi}_1^b\partial_b{\hat\xi}_2^a - {\hat\xi}_2^b\partial_b{\hat\xi}_1^a
+ g^{ab}\left( {\hat\xi}_1^t\partial_b{\hat\xi}_2^t -
{\hat\xi}_2^t\partial_b{\hat\xi}_1^t \right) .
\label{a13a}
\end{eqnarray}
The equality (\ref{a13}) will be used in the next section to compute the
central charge.  For now, let us note that when evaluated at a stationary
black hole metric in standard coordinates, for which $K_{rr} = 0 =
K_{\alpha\beta}$, the boundary term in (\ref{a11}) becomes
\begin{equation}
-{1\over8\pi G} \int_{r=r_+}\!\! d^{n-2}x\, \sqrt{\sigma}\left\{
{1\over f^2}\left[ \partial_r(f{\hat\xi}_2^r)\partial_r{\hat\xi}_1^t -
  \partial_r(f{\hat\xi}_1^r)\partial_r{\hat\xi}_2^t \right]
  + {1\over f}\partial_r\left[ {\hat\xi}_1^r\partial_r{\hat\xi}_2^t
  - \delta_{{\hat\xi}_2}{\hat\xi}_1^t \right]
   \right\} .
\label{a14}
\end{equation}
If we assume, as suggested by eqn.\ (\ref{a13a}), that
$\delta_{{\hat\xi}_2}{\hat\xi}_1^t = {\hat\xi}_2^a\partial_a{\hat\xi}_1^t$,
then this expression is antisymmetric in ${\hat\xi}_1$ and ${\hat\xi}_2$,
as required by eqn.\ (\ref{a13}).  Our boundary terms are thus consistent 
with the interpretation of $L[{\hat\xi}]$ as the generator of surface 
deformations in the presence of a horizon.

\section{The Virasoro Algebra}

In the preceding section, we considered general variations of a general
black-hole-like metric.  Let us now specialize to the case of an axially
symmetric black hole, with an adapted angular coordinate $\phi$ such
that $\partial_\phi g_{\mu\nu} = 0$.  For simplicity, I will assume
that only the component $N^\phi$ of the shift vector is nonzero; the
higher-dimensional generalization to more than one rotational Killing
vector is straightforward.

We now focus our attention on a particular subalgebra of the surface
deformation algebra with the following three properties:
\begin{enumerate}
\item The surface deformations are restricted to the $r$--$t$ plane.  This
specialization is inspired by the path integral approach to black hole
thermodynamics, in which it is clear that the $r$--$t$ plane has the central
role in determining the entropy \cite{Banados}.
\item The diffeomorphism parameter $\xi^t = {\hat\xi}^t/N$ ``lives on the
horizon,'' in the sense that near $r=r_+$ it depends on $t$ and $r$ only
in the combination $t-r_*$, where $fdr = Ndr_*$ in the time-slicing such
that $N_r=0$.  For the Kerr black hole, $t-r_*$ is essentially the standard
Eddington-Finkelstein retarded time, up to corrections of order $r-r_+$.
\item The lapse function $N^2$ is fixed at the horizon.  The horizon
is physically defined by the condition $N^2=0$, while our boundary term
(\ref{a10}) is written at $r=r_+$; this condition ensures that the boundary
remains at the horizon.
\end{enumerate}

Condition $1$ and eqn.\ (\ref{a6}) imply that the diffeomorphism parameter
$\xi^\phi$ has the form
\begin{equation}
\xi^\phi = -N^\phi\xi^t = -{N^\phi\over N}{\hat\xi}^t .
\label{b1}
\end{equation}
Condition $2$ requires that
\begin{equation}
\partial_r\xi^t = -{f\over N}\partial_t\xi^t
\label{b2}
\end{equation}
at $r=r_+$, allowing us to write radial derivatives at the horizon in terms
of time derivatives.  To impose condition $3$, we can examine diffeomorphisms
of $g^{tt} = -1/N^2$.  With initial coordinates chosen so that $N_r=0$, we
find that
\begin{equation}
\delta g^{tt} = 0
= {2\over N^2}\left( \partial_t - N^\phi\partial_\phi\right)\xi^t
+ {h\over N^4}\xi^r .
\label{b23}
\end{equation}
This structure suggests that we split our diffeomorphisms into left-moving 
modes $\xi^t$, for which $\partial_t\xi^t = \Omega\partial_\phi\xi^t$, and 
right-moving modes $\tilde\xi^t$, for which $\partial_t{\tilde\xi}^t = 
-\Omega\partial_\phi{\tilde\xi}^t$, where $\Omega = -N^\phi(r_+)$ is the 
angular velocity of the horizon.  Then
\begin{equation}
\xi^r = -{4 N^2\over h}\partial_t\xi^t , \qquad {\tilde\xi}^r = 0.
\label{b4}
\end{equation}
Note that $\xi^{r_*} = (f/N)\xi^r$ is, like $\xi^t$, a function of retarded 
time $t-r_*$ at the horizon.

We can use these results to write the left-moving modes at the horizon as
\begin{equation}
\xi^t_n = {T\over4\pi}\exp\left\{ {2\pi i n\over T}
\left( t - r_* + \Omega^{-1}\phi \right) \right\} ,
\label{b5}
\end{equation}
where $T$ is an arbitrary period.  (A possible choice is $T=\beta$, which 
matches the periodicity of the Euclidean black hole, but as we shall see, 
$T$ drops out of the final expression for the entropy.)  The normalization 
(\ref{b5}) has been fixed by the requirement that the surface deformation 
algebra (\ref{a13a}) reproduce the $\hbox{Diff}\,S^1$ algebra
\begin{equation}
\{ {\hat\xi}_m, {\hat\xi}_n \}_{\hbox{\scriptsize SD}}^t =
i(n-m){\hat\xi}_{m+n}^t .
\label{b6}
\end{equation}
Substituting the modes (\ref{b5}) into the boundary term (\ref{a14}), we
obtain
\begin{equation}
\delta_{{\hat\xi}_m} L[{\hat\xi}_n] = \hbox{\em bulk terms} +
  {A\over8\pi G}{\beta\over T}in^3\delta_{m+n} ,
\label{b8}
\end{equation}
where $A$ is the area of the boundary at $r=r_+$.  We can now use a trick 
of Brown and Henneaux to evaluate the central term $K[{\hat\xi}_m,
{\hat\xi}_n]$.  When evaluated on shell, the Hamiltonian and momentum 
constraints vanish, so $H[{\hat\xi}]=0$.  Equation (\ref{a13}) thus 
reduces to a collection of boundary terms,
\begin{equation}
{A\over8\pi G}{\beta\over T}in^3\delta_{m+n} =
J[\{ {\hat\xi}_m, {\hat\xi}_n \}_{\hbox{\scriptsize SD}}] +
K[{\hat\xi}_m,{\hat\xi}_n] =
i(n-m)J[{\hat\xi}_{m+n}] + K[{\hat\xi}_m,{\hat\xi}_n] .
\label{b9}
\end{equation}
{}From eqn.\ (\ref{a10}), it is easily checked that
\begin{equation}
J[{\hat\xi}_p] = {A\over16\pi G}{T\over\beta}\delta_{p0}
\label{b10}
\end{equation}
on shell.  Hence
\begin{equation}
K[{\hat\xi}_m,{\hat\xi}_n] = 
 {A\over8\pi G}{\beta\over T}in(n^2-{T^2\over\beta^2})\delta_{m+n} ,
\label{b11}
\end{equation}
the correct form\footnote{The $n$ dependence in (\ref{b11}) can be made 
the usual $n(n^2-1)$ by shifting $L_0$ by a constant.  This alters the 
eigenvalue (\ref{b10}), but also changes the ``effective central charge'' 
so that the entropy (\ref{c2}) is unaffected.} for the central term of 
a Virasoro algebra with central charge
\begin{equation}
c = {3A\over2\pi G}{\beta\over T} .
\label{b12}
\end{equation}

\section{Counting States}

The models we are investigating are not two-dimensional.  Nevertheless, 
the results above imply that the quantum states that characterize a black 
hole horizon must transform under a Virasoro algebra with central charge 
(\ref{b12}).  This is sufficient to permit the use of powerful methods 
from conformal field theory to count states.

In particular, a conformal field theory with a central charge $c$ has
a density of states $\rho(L_0)$ that grows asymptotically as
\begin{equation}
\log\rho(L_0) \sim 2\pi\sqrt{ {c_{\hbox{\scriptsize eff}}L_0\over6} } ,
\label{c1}
\end{equation}
where $c_{\hbox{\scriptsize eff}}$ is an ``effective central charge''
\cite{Cardy,Kutasov}.  If the spectrum satisfies reasonable, although not 
universal, conditions \cite{Carlip1}---notably that the ground state is an 
eigenstate of $L_0$ with eigenvalue zero---then $c_{\hbox{\scriptsize eff}} 
= c$.  Following Strominger \cite{Strominger2}, let us assume these 
conditions are satisfied in quantum gravity.  Then from eqns.\ (\ref{b10}) 
and (\ref{b12}), 
\begin{equation}
\log\rho(L_0) \sim {A\over4G} ,
\label{c2}
\end{equation}
recovering the standard Bekenstein-Hawking entropy.  In general, 
right-moving modes may make an additional contribution to the density
of states, but it is clear from eqn.\ (\ref{b4}) that the central charge
for those modes vanishes, so eqn.\ (\ref{c2}) gives the full entropy.

\section{Four Questions and Two Answers}

The analysis above strongly suggests that any quantum description of 
black hole horizon states must yield the standard Bekenstein-Hawking 
entropy.  Here, I will briefly address some details of this analysis and 
raise several remaining questions.\\[-.8ex]

\noindent {\bf 1.}\ What is the significance of the boundary condition
$N=0$?\\[-.8ex]

For a stationary black hole, in the coordinates $N^r=0$, $f\sim 1/N$ that we
used to evaluate the central charge, this is the condition for an apparent
horizon.  In other coordinates, however, the apparent horizon condition
is considerably more complicated.  An investigation of possible alternative
boundary conditions might help answer the question of what kind of ``horizon''
is needed for black hole entropy.  It may also be possible to extend this
analysis to more general gravitational actions along the lines of Ref.\
\cite{Brown3}.\\

\noindent {\bf 2.}\ The Cardy formula (\ref{c1}) comes from two-dimensional
conformal field theory.  What are the two relevant dimensions here?\\[-.8ex]

The Cardy formula requires a modular invariant partition function of the
form
\begin{equation}
Z = \hbox{Tr}\,\exp\{ i (J\phi + Et)\} ,
\label{d1}
\end{equation}
where $J$ and $E$ are conserved charges associated with translations in
$\phi$ and $t$.  For an axially symmetric black hole, $\phi$ and $t$ are
determined by the two Killing vectors, and modular invariance is a 
consequence of diffeomorphism invariance.  The two ``preferred''
directions are thus picked out by the symmetries.  (For a black hole in
more than four dimensions with more than one axial Killing vector, the
left-moving modes are determined by the condition $\partial_t\xi^t =
-N^\alpha\partial_\alpha\xi^t$, so the shift vector $N^\alpha$ picks
out an angular direction.)\\

\noindent {\bf 3.} What specific degrees of freedom account for the entropy
(\ref{c2})?\\[-.8ex]

Like Strominger's derivation \cite{Strominger2}, this computation does not
address this question, but rather uses symmetry arguments to derive the
behavior of any microscopic theory of black hole horizon states.  This is
both good and bad: good because it provides a universal explanation of black
hole statistical mechanics, bad because it offers little further insight
into quantum gravity.

One possible picture of the microscopic degrees of freedom comes from
considering the dimensional reduction of Einstein gravity to the $r$--$t$
plane near a horizon.  The resulting action contains a scalar field, 
essentially $\sqrt{\sigma}$, that couples to the two-dimensional scalar 
curvature ${}^{(2)}R$.  The action is not conformally invariant, but we 
know from the $c$ theorem \cite{Zam} that it must flow to a conformal 
field theory---presumably a Liouville theory---under the renormalization 
group.  Since the dimensionally reduced action has a prefactor of $A/16\pi G$, 
the central charge of such a Liouville theory is likely to be proportional 
to $A/16\pi G$, and might reproduce the central charge (\ref{b12}).  De 
Alwis has considered a similar renormalization group flow in a somewhat 
different context \cite{DeAlwis}, and Solodukhin has recently proposed a 
related analysis of dimensionally reduced gravity \cite{Solo}.

It also seems plausible that the description of black hole entropy here is
related to the picture of microscopic states as ``would-be pure gauge''
degrees of freedom that become dynamical at a boundary \cite{Carlip2,Bal}.
The existence of a central charge is an indication that the algebra of
surface deformations has become anomalous, and that invariance cannot be
consistently imposed on all states.  This connection has not been
developed, however.\\

\noindent {\bf 4.}\ Does the relevant conformal field theory satisfy the
technical conditions required for the Cardy formula?  In particular, does
the ground state have $L_0=0$?\\[-.8ex]

Without a much more detailed description of the conformal field theory, 
this question cannot be answered.  My approach differs from Strominger's 
in an important respect.  Strominger's boundary conditions were those of 
anti-de Sitter space, offering the possibility that anti-de Sitter space is 
the ground state.  The boundary conditions of this paper are those of a 
specific black hole, and depend on the horizon metric.  This difference 
is hard to avoid, since without the extra length scale provided by a 
cosmological constant it is difficult to write down a dimensionless central 
charge independent of the boundary values of the metric.  (One could, of
course, choose $T$ to be proportional to $A/\beta$ in eqn.\ (\ref{b12}), 
but there seems to be little physical justification for such a choice.)

There is, however, a plausible candidate for the ground state in the model
developed here: the extremal black hole, which is typically characterized 
by a lapse function behaving as $N^2\sim(r-r_+)^2$ near the horizon.  Such 
a configuration satisfies the boundary conditions assumed in this paper, but 
in contrast to the nonextremal result (\ref{b10}), eqn.\ (\ref{a10}) now 
gives $J[{\hat\xi}_0]=0$, implying that at least the classical contribution 
to $L_0$ vanishes.

\vspace{1.5ex}
\begin{flushleft}
\large\bf Acknowledgements
\end{flushleft}

I would like to thank Rob Meyers for pointing out some errors in the first
version of this paper.  This work was supported in part by National Science
Foundation grant PHY-93-57203 and Department of Energy grant DE-FG03-91ER40674.

\end{document}